**Band Structure and Spin-Orbital Texture of the (111)-KTaO$_3$ Two-Dimensional Electron Gas**

*Flavio Y. Bruno\*, Siobhan McKeown Walker, Sara Riccò, Alberto de la Torre, Zhiming Wang, Anna Tamai, Timur K. Kim, Moritz Hoesch, Mohammad S. Bahramy and Felix Baumberger*

Dr. F.Y. Bruno, Dr. S. McKeown Walker, Dr. S. Riccò, Dr. A. de la Torre, Dr. A. Tamai, and Prof. F. Baumberger
Department of Quantum Matter Physics, University of Geneva, 24 Quai Ernest-Ansermet,
1211 Geneva, Switzerland.
E-mail: flavio.bruno@unige.ch
Dr. Z. Wang, Prof. F. Baumberger
Swiss Light Source, Paul Scheerer Institute, 5232 Villigen, Switzerland.
Dr. Z. Wang
CAS Key Laboratory of Magnetic Materials and Devices, Ningbo Institute of Materials Technology and Engineering, Chinese Academy of Sciences, Ningbo, Zhejiang 315201, China
Dr. T.K. Kim, Dr. M. Hoesch
Diamond Light Source, Harwell campus, Didcot OX11 0DE, United Kingdom
Dr. M. Hoesch
Deutsches Elektronen-Synchrotron DESY, Photon Science, Hamburg 22607, Germany.
Dr. M. S. Bahramy
Quantum-Phase Electronics Center, Department of Applied Physics, University of Tokyo,
113-8656 Tokyo, Japan and RIKEN Center for Emergent Matter Science, 351-0198 Wako, Japan

Two-dimensional electron gases (2DEGs) in oxides show great potential for discovering new physical phenomena and at the same time hold promise for electronic applications. In this work we use angle resolved photoemission to determine the electronic structure of a 2DEG stabilized in the (111)-oriented surface of the strong spin orbit coupling material KTaO$_3$. Our measurements reveal multiple sub-bands that emerge as a consequence of quantum confinement and form a six-fold symmetric Fermi surface. This electronic structure is well reproduced by self-consistent tight-binding supercell calculations. Based on these calculations we determine the spin and orbital texture of the 2DEG. We show that the 2DEG Fermi surface is derived from bulk $J = 3/2$ states and exhibits an unconventional anisotropic Rashba-like lifting of the spin-degeneracy. Spin-momentum locking holds only for high symmetry directions and a strong out-of-plane spin component renders the spin-texture three-fold symmetric. We find that the average spin-splitting on the Fermi surface is an order of magnitude larger than in $SrTiO_3$, which should translate into an enhancement in the spin-orbitronic response of (111)-$KTaO_3$ 2DEG based devices.

The realization of two dimensional electron gases (2DEGs) with remarkable physical properties at complex oxide interfaces has become one of the major driving forces in the field of oxide electronics.[1,2] The correlated materials used in these heterostructures display fascinating bulk properties, ranging from high temperature superconductivity to ferroelectricity and Mott metal-insulator transitions, among others. It is this plethora of bulk properties that enrich the phase diagrams of transition metal oxides compared to those of conventional semiconductors, and which renders 2DEGs in correlated materials so promising for the discovery of new functionalities and emergent physical phenomena. This is exemplified by the observation of superconductivity[3], magnetism[4], strong electron lattice interactions[5] and quasi 1D conductivity[6] in the 2DEG stabilized in (001)-oriented $SrTiO_3$ (STO) which is, by far, the most studied system among $ABO_3$ transition metal oxides. At present it is possible to create confined electron systems in other several oxides including: $KTaO_3$, $BaTiO_3$ and $TiO_2$ by different synthesis strategies.[7–10]. While the various methods used to engineer such 2DEGs may differ, the physics of the resulting system is always dominated by the host crystal. This is an advantageous situation since different methods allow 2DEGs to be obtained in diverse forms, thereby expanding the range of experimental probes to which these systems are accessible. A significant step forward in the understanding of these systems came with the experimental observation of the electronic structure of 2DEGs by angle resolved photoemission spectroscopy (ARPES).[11–16] This facilitated direct comparisons with electronic structure calculations that have provided the foundations for the understanding of the electronic properties of these systems.

The lifting of spin degeneracy due to an inversion-symmetry breaking electric field normal to a surface and/or heterointerface is broadly known as the Rashba effect. Its presence in oxide 2DEGs has been inferred from magnetotransport experiments and theoretically investigated in (001)-oriented STO and KTO 2DEGs.[17–20] Very recently there has been a surge in attention directed towards understanding the spin-orbit coupling mechanism, and the spin texture of oxide 2DEGs in general, due to their potential in spin-orbitronics applications.[21–24] Interest was further raised by the observation of spin-to-charge conversion at the (001)-LAO/STO interface with a figure of merit that far exceeded the record held by traditional spintronic materials[25,26]. For a single parabolic band with Rashba spin splitting, the figure of merit used to quantify the conversion efficiency of 3D spin to 2D charge current is proportional to the Rashba coupling constant $α_R$. This parameter describes the linear relationship between spin-up/-down energy splitting $Δ_R$ and the in-plane momentum ($k_∥$) in a free electron Rashba system according to $Δ_R = 2α_R k_∥$ [27]. While there is some evidence that this relationship is valid in STO 2DEGs at low carrier densities, the form of the Rashba spin-splitting in quantum confined multi-orbital systems can be significantly more complex.[18,19,23]. When looking for spintronic systems a common design rule is to maximize $α_R$ of the 2DEG, and thus the magnitude of Rashba spin-splitting, through the choice of parent material.[20] This is done by looking for materials with large atomic spin orbit coupling ($\xi$) and by looking for ways to break inversion symmetry at a microscopic level, which are both thought to augment $\alpha_R$. Following these classic design rules, KTO appears to be an attractive analog to STO for oxide spin-orbitronics. As well as being isostructural to STO, the spin-orbit coupling (SOC) strength is $\xi_{KTO} \approx 0.4$ eV for the Ta $5d$ electrons, while $\xi_{STO} \approx 0.02$ eV for the Ti $3d$ electrons in STO.

The bulk band structure of KTO and STO is shown in **Figure 1** where the different effects of SOC in KTO and STO are evident. The conduction band minimum in KTO is at the Γ point and consists of two spin-degenerate $J = 3/2$ bands. The members of this doublet have different band masses, evidenced from the light and heavy bands observed in transport experiments.[28] Due to the large spin orbit interaction the $J = 1/2$ band sits 400 meV above and does not contribute to the electronic conduction. In the case of oxide 2DEGs, the roles played by multi-orbital physics, the strong crystal field, spin-orbit coupling, quantum confinement and their interplay, in defining the electronic structure in general and subsequently the Rashba coupling constant $\alpha_R$, should be carefully examined. To date, most studies have focused on the (001)-oriented 2DEG systems. A combined theoretical and experimental exploration of 2DEGs in different orientations offers the attractive possibility of defining the design rules in more detail.[29,30]. Interest in (111) oriented oxide heterostructures has surged recently,[31–33] and is reinforced by theoretical proposals for topological states in (111) oxide based 2DEGs.[34–36] Furthermore, considering the recent surprising results in STO based systems further progress in the understanding of KTO 2DEGs holds a great potential for spin-orbitronic functionalities.

In this communication we present ARPES measurements of the electronic structure of a 2DEG confined in the (111)-KTO surface. We thereby tune, both SOC and crystallographic orientation simultaneously, compared to the well-studied STO (001) system. Our measurements reveal a six-fold symmetric Fermi surface that is well reproduced by tight binding supercell calculations based on relativistic density functional calculations of the bulk electronic structure. Based on this band structure we study the spin texture formed as a consequence of quantum confinement and Rashba-like lifting of degeneracy in this system. We put our results into a broader context by comparing the band structure calculations to those of the (111)-STO 2DEG. As a measure for the spin-orbitronic response of (111)-KTO 2DEGs we calculate the difference in Luttinger volume for spin-split pairs of a Fermi surface sheet. We found this quantity in KTO to be an order of magnitude larger than in STO for the largest Fermi surface sheet. Additionally, we found that the spin texture of the (111)-KTO 2DEG differs from the classical Rashba picture in several ways: it has a sizeable out of plane spin component with three fold symmetry, the momentum splitting on different parts of the Fermi surface is neither isotropic nor a simple function of momentum, and spin-momentum locking holds only for high symmetry directions. We rationalize these findings in terms of the multi-orbital physics of these systems, bulk SOC, and the important role of the bulk crystal field.

We stabilized a 2DEG by creating oxygen vacancies in the surface of commercial (111)-KTO single crystals by irradiation with synchrotron light in UHV. Mobile carriers released by the oxygen vacancies then create a confined electronic system near the surface. All the measurements presented here were obtained after exposing the substrate to synchrotron light for a sufficiently long time to saturate the electron density (see Figure S1).[37,38] Our improved preparation method of the KTO surface allowed us to obtain measurements of the (111)-KTO 2DEG of unprecedented quality. The Fermi surface of the resulting 2DEG is obtained by ARPES and is shown in **Figure 2a**. It consists of a six-fold symmetric star-shaped contour centered at the Γ point with a second hexagonal contour within it. The star-shaped Fermi surface sheet has a major diameter of ~ 0.6 Å$^{-1}$ along the Γ-M direction and the minor diameter along Γ-K of ~ 0.3 Å$^{-1}$. The dispersion of the bands along these high symmetry directions is shown in Figures 2b and 2f. In Figure 2b we see two electron-like bands with approximately 130 meV bandwidth along the Γ-M direction. These bands have Fermi wavevectors $k_F$ = 0.26 and 0.14 Å$^{-1}$ and belong to the star-like and hexagonal Fermi surface contours respectively. They are clearly resolved as maxima in the momentum distribution curve (MDC) at $E_F$ shown in the inset. Along the Γ-K direction the same two bands overlap, and appear as a single contour in Figure 2f, which corresponds to where the star-like and hexagonal contours touch. Increased spectral weight near $E_F$ and $k_\parallel$ = 0 in the dispersion plots of Figure 2b and 2f reveal the existence of another electron like band with low density. These bands can be seen more clearly in the energy distribution curves (EDCs) at $k_\parallel$ = 0 shown in Figures 2c and 2g appearing as peaks in intensity at approximately 30 meV. From the simple argument that the bandwidth of the system is much less than the $\Delta_{SO}$ ~ 400 meV of bulk KTO (cf. Figure 1), and the energetic separation of the low and high density bands is only 100 meV, the low-density band cannot be attributed to the $J$ = 1/2 singlet. Rather, it is a higher order subband; a manifestation of the quantum confinement of electrons near the surface, and evidence for the two-dimensionality of this system. Curvature plots obtained from the dispersion plots further confirm the presence of these subbands (see Figure S2). In order to estimate the density of this system it is necessary to understand the nature of these bands, in particular their spin degeneracy.

With the aim of better understanding the electronic structure observed in our ARPES experiments we performed tight binding supercell calculations based on a relativistic *ab initio* bulk band structure. Bulk truncation and a potential well were imposed to emulate the band bending that arises at the surface due to positively charged oxygen vacancies. A self-consistent solution of the Poisson and Schrödinger equations, which are coupled by this spatially dependent potential, was found numerically. The magnitude of the potential at the surface was chosen to ensure the calculated band structure has a bandwidth comparable to what is observed by ARPES. The calculated band dispersions along Γ-M and Γ-K are shown in Figures 2d and 2h respectively and the calculated Fermi surface is shown in Figure 2e. The star-like and hexagonal Fermi surface contours are found, in good agreement with our data. The binding energies of the higher order subbands are also well reproduced by our calculation. This striking agreement without the need to include any *ad hoc* hypotheses encourages us to look more closely at the calculation to gain insight into the nature of the KTO 2DEG. To clarify the complementary roles of quantum confinement and atomic spin orbit coupling in defining the subband structure, spin texture and spin-splitting in oxide 2DEGs we also perform a comparative analysis of our (111)-KTO tight binding supercell calculations with an equivalent calculation for the (111)-STO 2DEG.

Our calculation confirms that the $J = 1/2$ band does not contribute to the (111)-KTO 2DEG since the bulk SOC splitting of 400 meV is preserved when the confinement is along the (111) orientation. It follows that all the observed subbands are derived from the $J = 3/2$ doublet. This is different to the case of (111)-STO where the subbands originating from both the $J = 1/2$ and the $J = 3/2$ band intersect each other, leading to a rather more complex band structure. In the bulk, while SOC causes significant orbital mixing at the Gamma point, at high momenta the $J = 3/2$ bands retain an almost singular orbital character as observed in Figure 1a. The color scale of the calculated subband structure indicates the $t_{2g}$ orbital content of the states in Figure 2d. From this we see that the singular orbital character of the bands is preserved in the 2DEG. The unique orbital character of each of the extremal parts of the star-like Fermi surface contour along the Γ-M directions is also evident from their red, blue or green color in Figure 2e. This corresponds to a dominant $d_{xy}$, $d_{yz}$ or $d_{xz}$ character, and demonstrates that each Fermi surface contour has contributions from all three $t_{2g}$ orbitals. Spatial confinement along the (111) direction has an equivalent effect on all $t_{2g}$ orbitals and as a consequence, neither (111)-STO nor KTO 2DEGs show dramatic orbital polarization in momentum-integrated measurements.[39,40] This is demonstrated by the common bandwidth of all the extrema of the star-like Fermi surface, which is predicted by our calculations, and confirmed by our measurements. However, as seen near $k_\parallel = 0$ in Figure 2d and 2h, our calculation suggests that the degeneracy of the $J = 3/2$ bands at the Gamma point is lifted. This can be attributed to the lower symmetry of the trigonal crystal field at the (111) surface which lifts the degeneracy of the $e_g$' manifold [34], though the splitting is not resolved by our experiment.

An important aspect of these calculations is that, by including the potential well at the surface, the 2DEG and Rashba-type spin-splitting emerge without the need to explicitly including symmetry breaking terms by hand. This follows from the long-range hopping terms and spin-orbit coupling included in our realistic tight binding model derived from density functional theory calculations of bulk KTO. Therefore, the supercell calculation provides direct insight into the 2DEG spin texture. While the spin splitting that we describe here is not observed in our experiment, since it is of such paramount importance in spintronic experiments, we will explore it in more detail. **Figure 3a** and 3b show the first four subbands in the Fermi surface for (111)-STO and KTO 2DEGs respectively. In KTO we see that the spin degeneracy of both the star-like and hexagonal Fermi surface contour is lifted, which allows us to estimate the density of the state observed experimentally. From the area of the star-like and hexagon-like Fermi surface sheets we estimate a 2D electron density of $n_{2d} = 1.2 \times 10^{14}$ cm$^{-2}$. This is a lower limit of the 2DEG density since we only consider the filling of the lowest order subbands. This carrier density is similar to that observed in 2DEGs on (001)-KTO by previous ARPES experiments suggesting that the creation of oxygen vacancies by photons, or the screening mechanism, does not depend strongly on surface orientation. This density is also similar to that obtained in similar experiments on STO and displays the same saturating behavior.[13,14]

In the free electron 2DEG model proposed by Bychkov and Rashba, the two concentric circles that form the spin-split Fermi surface have in-plane, chiral spin windings of opposite sign for each contour.[41] We observe the same characteristic here as shown by the arrow vectors at certain points of the star-like Fermi surface contours in Figure 3 where the color scale represents inner (dark) and outer (light) contour of a spin-split pair and the arrows represent the in-plane spin vector. This spin momentum locking holds only along the Γ-M and Γ-K directions for all subbands. However, in contrast to the classic Rashba picture, the momentum splitting on different parts of the Fermi surface is not constant, to an extent that cannot be explained by the expected proportionality to $k_{\parallel}$. To quantify this, we can use the definition of $\alpha_R$ from the free-electron Rashba picture to define an effective Rashba constant at different points on the Fermi surface. This effective $\alpha_R$ will have a strong $k_{\parallel}$ dependence as observed in Figure 3 and can be used as a comparative measure of Rashba like spin-splitting. Our calculation results in $\alpha_{R\Gamma M}^{KTO}$ = 2 meV Å at the tip of the star, and a momentum splitting $\Delta k_{\Gamma M}^{KTO} \sim 2 \cdot 10^{-3}$ Å$^{-1}$ along the Γ-M direction as shown in Figure 3c. In Figure 3d the splitting along the Γ-K direction is $\Delta k_{\Gamma K}^{KTO} \sim$ 5·10$^{-3}$ Å$^{-1}$ obtaining $α_R^{KTO}$ = 21 meV Å at the intersection of the star's lobes where the Fermi contour changes orbital character. This indicates that the effective strength of Rashba-like spin orbit coupling is higher at these intersections. This is reminiscent of the enhanced Rashba spin-splitting predicted at the avoided crossings of light and heavy bands in (001)-STO 2DEGs. Indeed, our (111)-KTO calculation reveals that the orbital angular momentum (OAM) is enhanced in the regions of these avoided crossings, as would be expected within the framework of a OAM induced Rashba-like spin splitting[42]. These intersections can also be described as avoided crossings. In contrast to the case of (001)-STO where the avoided crossings only appear as the heavy $d_{xz/yz}$ subbands become significantly populated, the avoided crossing to the (111)-STO and KTO 2DEGs extend over the full bandwidth of the system, as is indicated by the dashed box in Figure 2h. Interestingly, the strength of OAM and the effective $α_R^{KTO}$ value at this avoided crossing have an energy dependence, indicating that some modulation of SOC in both (111)-STO and KTO 2DEGs would be expected and should be strongest for low densities, as was recently observed in magnetotransport experiments.[43]

Due to the reduced SOC in bulk STO relative to KTO the naïve expectation is that Rashba-like spin orbit coupling should be an order of magnitude smaller in STO. In our calculation for (111)-STO we find that along the Γ-K direction $Δk_{ΓK}^{STO}$ ~ 2·10$^{-3}$ Å$^{-1}$ ($α_{RΓK}^{STO}$ = 5 meV Å) at the Fermi level as shown in figure 3f. This gives the somewhat counter intuitive result that the maximum Rashba splitting $Δk$ in STO and KTO is only different by a factor of ~ 3 (factor of ~ 4 in $α_R$). However, at the Fermi level in the Γ-M direction, we show a value of $Δk_{ΓM}^{STO}$ ~ 1·10$^{-4}$ Å$^{-1}$ and a corresponding $α_{RΓM}^{STO}$ = 0.05 meV Å as depicted in Figure 3e. This is indeed approximately a factor of 40 smaller than the equivalent value in KTO. This suggests that at the avoided crossings, while SOC surely play some role, the magnitude of the maximum Rashba spin splitting is controlled by the enhancement of OAM, which is more dramatic in STO than KTO. In any spin-orbitronic application, the contribution of the entire Fermi surface should be considered. To this end, we calculate the change in Luttinger volume of the largest Fermi surface sheet in STO and KTO, normalized by the carrier density. We find that in KTO there is a 1% change in volume while in STO there is only a 0.1% change. It can be attributed to the significant spin-splitting found everywhere on the Fermi surface in the (111)-KTO 2DEG unlike for STO based 2DEGs and to the fact that the large splitting along Γ-K is more spread in the former than in the latter (*cf*. Figures 3d and 3f). This should translate into a quite dramatic enhancement in the spin-orbitronic response of (111)-KTO 2DEG based devices.

In **Figure 4** we show the spin texture for the split first subband in a (111)-KTO 2DEG where further differences compared to the conventional Rashba picture are clear. The arrows indicate the in-plane spin direction and magnitude while its color indicates the out-of-plane component. Away from the high symmetry directions the spins of a spin-split pair wobble in opposition around the Fermi surface. Interestingly, for confinement of (111)-KTO and STO along the direction, our calculations predict a sizeable out of plane spin expectation value. In contrast to the 6-fold symmetry of the electronic dispersion, the out-of-plane spin component presents a three-fold symmetry. This characteristic allows the detection of the out-of-plane spin texture in magneto-transport experiments, as was recently demonstrated in (111)-STO 2DEGs.[44] That our calculation shows a similar out of plane spin component in both of these systems, provides further evidence that it is a general feature of perovskite based 2DEGs confined in the (111) plane. It is a manifestation of the principle that the preferential direction of polarization is imposed by the crystal field, and plays an important role in defining the spin structure of an electronic system.

In summary, we have measured the electronic band structure of the 2DEG stabilized in the surface of (111)-KTO by means of ARPES. We resolved a Fermi surface formed by a six-fold symmetric star-shaped contour with a second hexagonal contour within it. The experimentally determined electronic structure is well described by tight-binding supercell calculations. We demonstrated that the 2DEG is formed by bands derived from the $J = 3/2$ doublet and the Fermi surface contours have contributions from all three $t_{2g}$ orbitals. Our calculation predicts a sizable momentum dependent Rashba splitting of the Fermi surface with a complex spin-texture presenting a substantial out-of-plane spin component and wobbling of the in-plane spin component. Based on our results we suggest that the maximal spin-splitting at the Fermi level is not a good metric for the performance of a material for spintronic applications. Indeed, despite the only moderate enhancement of the maximal $α_R$, an analysis of Luttinger volume suggests that -(111)-KTO is a very good candidate because the large bulk SOC in KTO leads to significant spin-splitting everywhere on the Fermi surface unlike STO based 2DEGs. Together with the predicted novel out of plane spin component with three-fold symmetry in the (111)-KTO 2DEG makes the material an ideal candidate to explore new possibilities in spin-orbitronic devices.

**Experimental Section**

**ARPES measurements:** The 2DEG was stabilized in the surface of commercial $KTaO_3$ (111) single crystals. As-received crystals were annealed for 30 minutes at 450°C in UHV (P<$10^{-9}$ mbar). After annealing, a small dose of K was evaporated onto the surface which helps to prevent charging effects during the ARPES measurements that arise due to the insulating nature of the substrate. It is important to note that the 2DEG is induced by oxygen vacancies created as a consequence of light irradiation and is not dependent on the amount of K deposited (see Figure S1). ARPES measurements were taken at T=10 K at the I05 ARPES beamline of the Diamond Light Source using a photon energy of 96 eV.[45]

**Calculations:** The relativistic electronic structure calculations were performed within the density functional theory (DFT) using the Perdew-Burke-Ernzerhof (PBE) exchange-correlation functional, as implemented in the WIEN2k package[46]. A 10×10×10 *k*-mesh was used to sample the Brillouin zone and the muffin-tin radius $R_{MT}$ for all atoms was chosen such that its product with the maximum modulus of reciprocal vectors $K_{max}$ becomes $R_{MT}K_{max}$ = 7.0. To describe the quantum confinement at the surface, the bulk DFT calculations were downfolded using maximally localized Wannier functions[47,48] made of Ta-$t_{2g}$ orbitals, and the resulting 6-band tight-binding transfer integrals implemented within a 30-unit (90 Ta Layers) supercell, with an additional on-site potential term, accounting for the electrostatic band bending potential. We solved this self-consistently with Poisson's equation assuming an electric-field dependent dielectric constant modeled according to reference[49]. The only adjustable parameter is the potential that is varied until the experimental bandwidth is reproduced.

**Supporting Information**

Supporting Information is available from the Wiley Online Library or from the author.

**Acknowledgements**

We gratefully acknowledge discussions with S. Okamoto and N. Sivadas. F. Y. B. was supported by SNSF Ambizione Grant No. PZ00P2-161327. M. S. B. gratefully acknowledges the financial supports from CREST, JST (Project No. JP-MJCR16F1). S. M. W., S. R., Z. W., A. d. l. T., A. T. and F. B. acknowledge the financial support by SNSF projects N° 146995 and 165791. We acknowledge Diamond Light Source for time on beamline I05 under proposal SI11741. We acknowledge the Swiss Light Source for time on SIS beamline under proposal 20150450.

Received: ((will be filled in by the editorial staff))
Revised: ((will be filled in by the editorial staff))
Published online: ((will be filled in by the editorial staff))

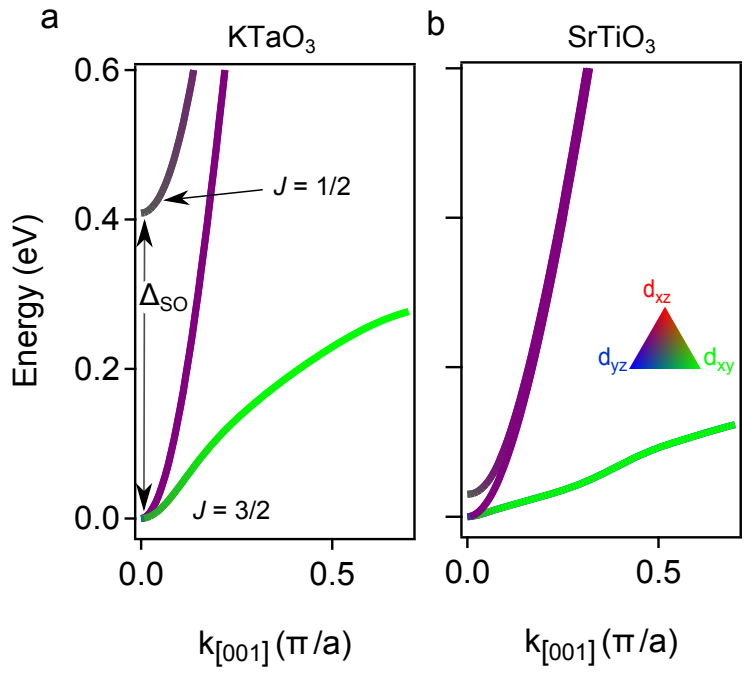

**Figure 1.** Bulk electronic structure and orbital character of (a) $KTaO_3$ and (b) $SrTiO_3$.

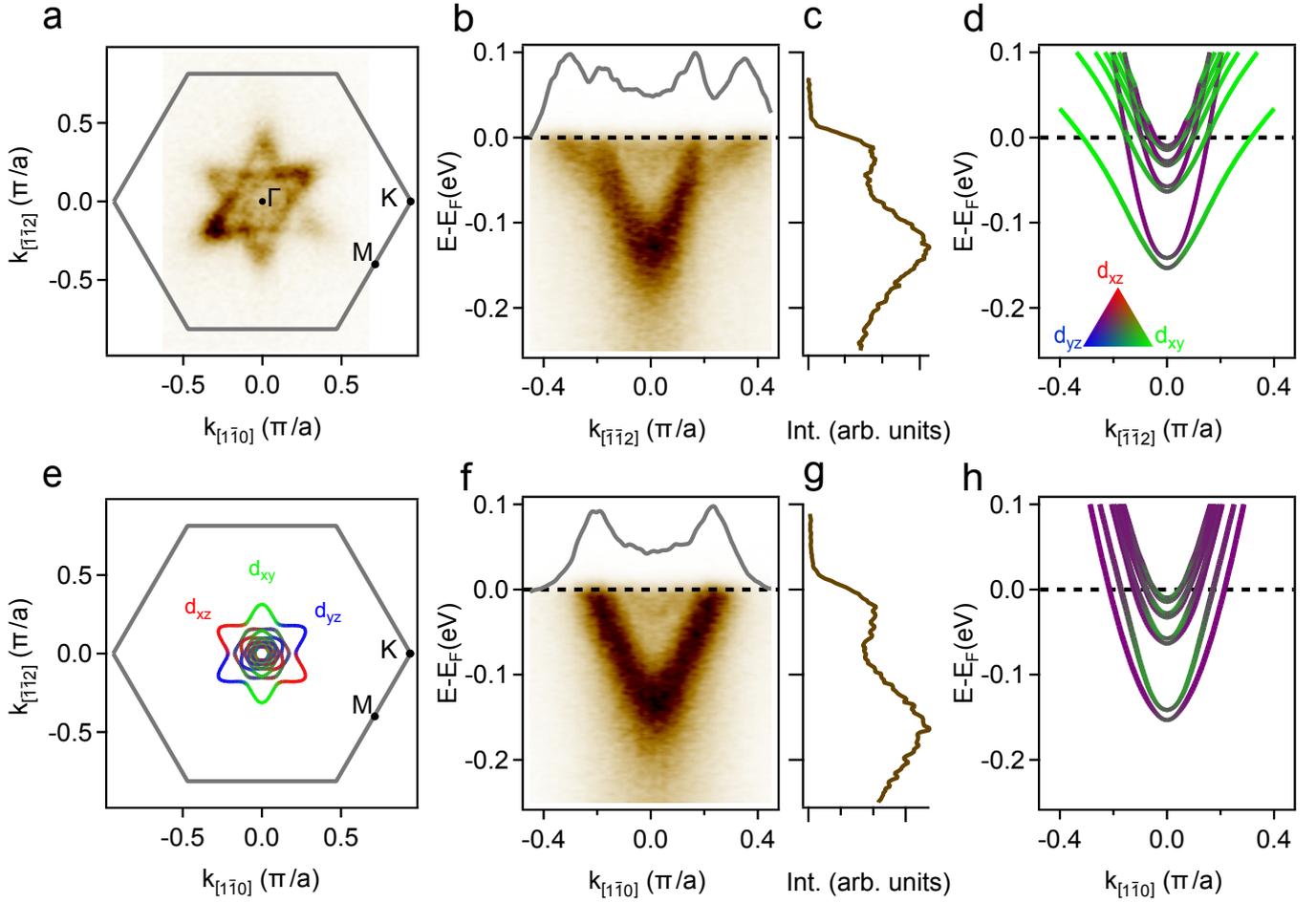

**Figure 2.** (a) Fermi surface of $KTaO_3$ two dimensional electron gas measured at 108 eV with LH polarized light. (b)-(f) Energy-momentum dispersion measured along high symmetry directions Γ-M and Γ-K respectively. Top inset momentum distribution curves at the Fermi level. (c)-(g) energy distribution curves at $k_{//}=0$. Tight-binding supercell calculations of the electronic structure. (e) Fermi surface with orbital character (d)-(h) calculated band dispersion along the high symmetry directions.

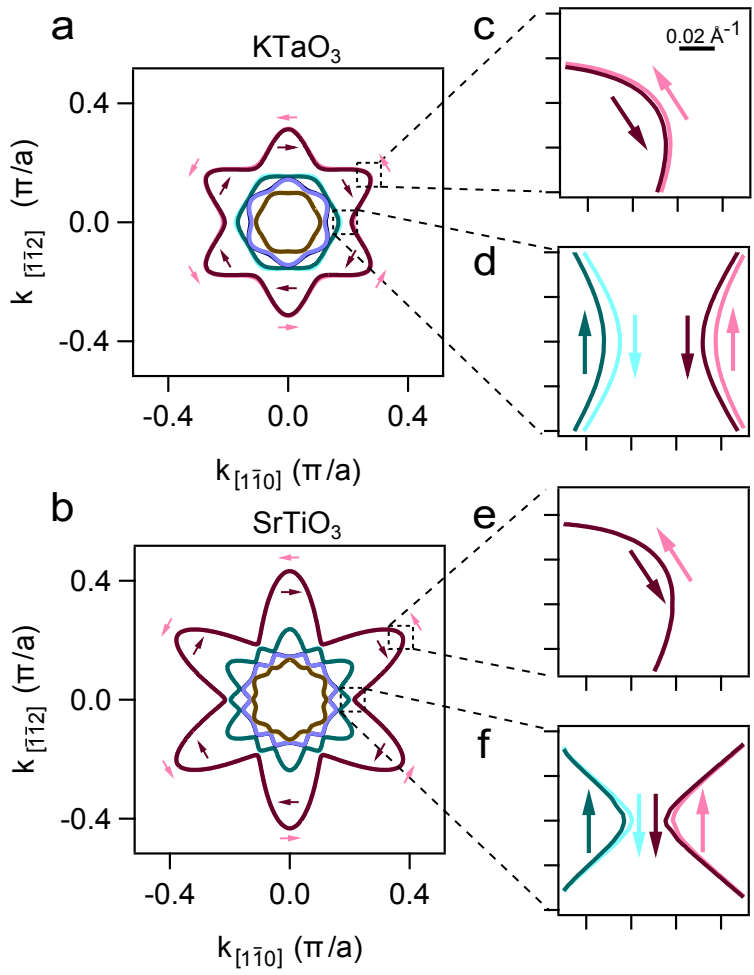

**Figure 3.** Tight-binding supercell calculations of the first 4 sub-bands in Fermi surface of (a) KTaO$_3$ and (b) SrTiO$_3$ two dimensional electron gas. The arrows indicate the spin direction for the Rashba-like split Fermi contour. (c)-(f) Enlarged view of the band splitting.

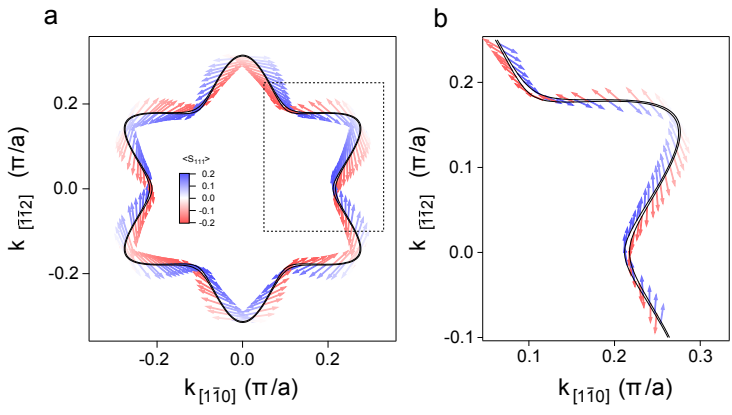

**Figure 4.** Calculated spin texture of the split first sub-band of a (111) - KTaO$_3$ two dimensional electron gas. The vector arrow denotes the in plane component and the color scale indicates the out-of-plane spin component.

**Two-dimensional electron gases (2DEGs) in oxides** show great potential for discovering new physical phenomena. The electronic structure of a 2DEG stabilized in the (111)-oriented surface of the strong spin orbit coupling material $KTaO_3$ is determined by means of angle resolved photoemission spectroscopy. This electronic system holds promise for spin-orbitronics based on the unconventional anisotropic Rashba-like lifting of the spin-degeneracy found in the system.

**Keyword: oxide electronics**

F. Y. Bruno*, S. McKeown Walker, S. Riccò, A. de la Torre, Z. Wang, A. Tamai, T. K. Kim, M. Hoesch, M. S. Bahramy and F. Baumberger.

**Band Structure and Spin-Orbital Texture of the (111)-$KTaO_3$ Two-Dimensional Electron Gas**

ToC figure

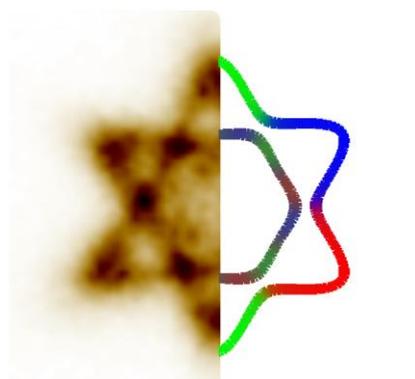

## Supporting Information

**Band Structure and Spin-Orbital Texture of the (111)-KTaO$_3$ Two-Dimensional Electron Gas**

*Flavio Y. Bruno\*, Siobhan McKeown Walker, Sara Riccò, Alberto de la Torre, Zhiming Wang, Anna Tamai, Timur K. Kim, Moritz Hoesch, Mohammad S. Bahramy and Felix Baumberger*

**1 Electron Density Evolution.**

Commercial single KTaO$_3$ crystals were annealed in a preparation chamber for 30 minutes at 450°C in a pressure P < 10$^{-9}$ mbar in order to clean the surface of contaminants. Subsequently, a small dose of K was evaporated on the surface under UHV conditions to achieve sufficient surface conductivity for the ARPES experiments. The pristine crystals used in our experiments are insulating and preclude obtaining high quality ARPES data due to accumulation of charge in the sample during measurements. We found that depositing a small amount of K is sufficient to create a lightly conducting surface layer enabling the grounding of the sample. Following these steps we transferred the sample to the measurement chamber were the photoemission experiments were carried out.

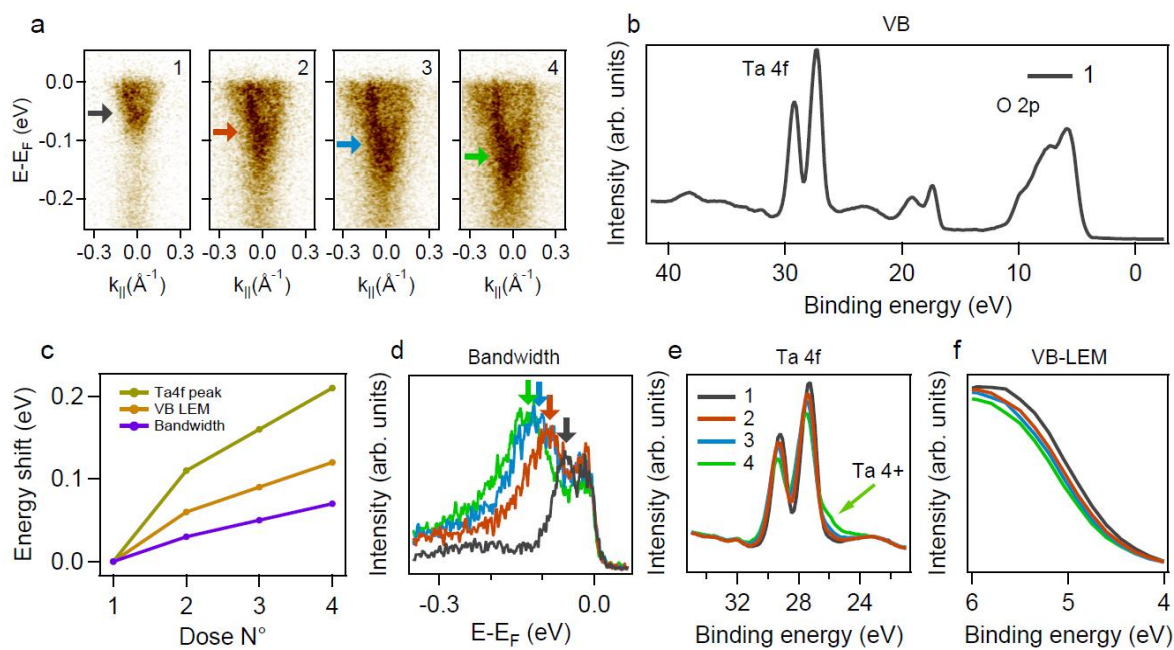

**Figure S1.** (a) Evolution of the electronic band dispersion of a (111) - KTaO$_3$ 2DEG with irradiation time. Labels 1-4 correspond to irradiation times of ~ 1, 2, 4, and 30 minutes with syncrotron light of 96 eV. (b) Integrated spectrum of the valence band and shallow core levels. (c) Energy shift of different spectral signatures with irradiation time. (d-f) Zoomed in view of the evolution of the integrated spectra with irradiation times showing details of the (d) bandwidth, (e) Ta 4f peak, and (f) valence band leading edge.

In Figure S1a we show the evolution of the 2DEG band dispersion with irradiation by 96 eV synchrotron light corresponding to exposure times of 1, 2, 4 and 30 minutes, we label these as states 1, 2, 3 and 4 respectively. Light irradiation creates positively charged oxygen vacancies that are localized in the surface at low temperatures. Mobile carriers will then screen these defects producing band bending as a consequence. The signatures of this process are illustrated in Figure S1. As already pointed out, the bandwidth of the system increases with increasing exposure time. This is again observed in the angle integrated spectra shown in figure S1d where the band bottom of the 2DEG is indicated by an arrow. In figure S1e we observe a shoulder in the Ta 4f peak that corresponds to $Ta^{4+}$ increasing in intensity with irradiation time, which is the signature of oxygen vacancies being created in the system. Finally, the valence band leading edge midpoint (VB-LEM) and the Ta 4f peak shifts towards higher binding energies as observed in Figures S1 e and f. All these effects are summarized in Figure S1c where we plot the energy shift of the electron band bottom, the VB-LEM and the Ta 4f peak referenced to the initial state. The energy shift is a consequence of the creation of oxygen vacancies at the surface and the concomitant band bending. [1] Importantly, the system becomes stable after ~ 20 minutes of irradiation and the bandwidth and electron density saturate to the maximal observed value. All measurements reported in the main manuscript were done on samples with saturated electron density.

## 2 Curvature Plots.

In Figure 2b and 2f of the main text we show the band dispersion in high symmetry directions. In Figure S2 a and b we show curvature plots of these data to better visualize the first and second sub-band.

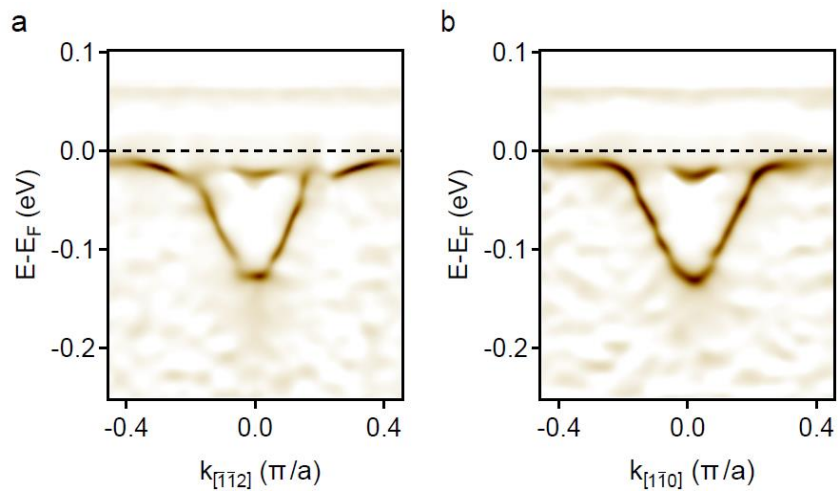

**Figure S2.** Curvature plots obtained from the energy-momentum dispersion plots of a (111)-$KTaO_3$ 2DEG along the high symmetry directions (a) Γ-M and (b) Γ-K.

**3 Photon Energy Dependence.**

In Figure S3 we show a Fermi surface map in the $k_z – k_{//}$ plane obtained by varying the photon energy between 65 and 125 eV. The parallel momentum $k_{//}$ is oriented along the Γ-M direction. Dashed lines indicate the Fermi wave vector of the two electron-like bands observed in this high symmetry direction (*cf.* Figure 2b). The perpendicular wave vector $k_z$ was calculated assuming free electron final states and an inner potential of 12 eV. We indicate with a blue circle the expected position for the Γ point in a bulk doped sample with 3D electronic structure.

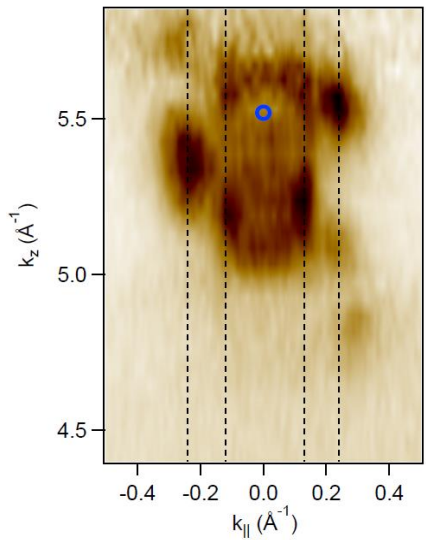

Figure S3. Fermi surface of the (111)- $KTaO_3$ 2DEG measured in the Γ-M direction with photon energies between 65 and 125 eV. The dashed lines indicate the Fermi wavevector of the electronlike bands. The blue circle indicates the position of the bulk Γ point.